\def\bq{\begin{equation}}
\def\eq{\end{equation}}
\def\bqa{\begin{eqnarray}}
\def\eqa{\end{eqnarray}}
\def\bqb{\begin{eqnarray*}}
\def\eqb{\end{eqnarray*}}
\documentstyle[12pt]{article}\textwidth 16cm
\def\pr#1#2#3{ Phys. Rev. ${\bf{#1}}$ (#2) #3}

\def\pl#1#2#3{ Phys. Lett. ${\bf{#1}}$ (#2) #3 }

\def\np#1#2#3{ Nucl. Phys. ${\bf{#1}}$ (#2) #3}
\def\zp#1#2#3{ Z. Phys. ${\bf{#1}}$ (#2) #3}

\global\nulldelimiterspace = 0pt

\def\Bsl{\hbox{/\kern-.6700em$B$}} 
\def\Dsl{\hbox{/\kern-.6700em$D$}} 
\def\Wsl{\hbox{/\kern-.6700em$W$}} 

\def\roughly#1{\mathrel{\raise.3ex
    \hbox{$#1$\kern-.75em\lower1ex\hbox{$\sim$}}}}

\def\A{ {\cal A }}

\def\mh2{m^2_H}

\begin{document}
\pagenumbering{arabic}
\thispagestyle{empty}
\hspace {-0.8cm} hep-ph/9509216\\
\hspace {-0.8cm} PM/95-35 \\
\hspace {-0.8cm} July 1995\\
\hspace {-0.8cm} corrected version\\
\vspace {0.8cm}\\

\begin{center}
{\Large\bf Universal and Non Universal New Physics Effects in a General
Four-fermion Process: a Z-peak Subtracted Approach} \\

 \vspace{1.8cm}
{\large  F.M. Renard$^a$ and C.
Verzegnassi$^b$}
\vspace {1cm}  \\
$^a$Physique
Math\'{e}matique et Th\'{e}orique,
CNRS-URA 768,\\
Universit\'{e} de Montpellier II,
 F-34095 Montpellier Cedex 5.\\
\vspace{0.2cm}
$^b$ Dipartimento di Fisica,
Universit\`{a} di Lecce \\
CP193 Via Arnesano, I-73100 Lecce, \\
and INFN, Sezione di Lecce, Italy.\\

\vspace{1.5cm}

 {\bf Abstract}
\end{center}
\noindent
We calculate, using a Z-peak subtracted representation of four-fermion
processes previously illustrated for the case of electron-positron
annihilation into charged lepton-antilepton, the corresponding
expressions of the new physics contributions for the case of final
quark-antiquark states, allowing the possibility of both universal and
non universal effects. We show that, in each case, the main result
obtained for the final lepton channel can be generalized, so that every
experimentally measurable quantity can be expressed in terms of input
parameters \underline{measured} on Z resonance, of $\alpha(0)$ and of a
small number of subtracted one loop expressions.
Some examples of models are considered for several
c.m. energy values, showing that
remarkable simplifications are often introduced by our approach. In
particular, for the case of a dimension-six lagrangian with anomalous
gauge couplings, the same reduced number of parameters that would
affect the observables of final leptonic states are essentially
retained when one moves to final hadronic states. This leads to great
simplifications in the elaboration of constraints and, as a gratifying
byproduct, to the possibility of making the signal from these models
clearly distinguishable from those from other (both universal and non
universal) competitors.

\vspace{1cm}

\setcounter{page}{0}
\def\thefootnote{\arabic{footnote}}
\setcounter{footnote}{0}
\clearpage

\section{Introduction}

At the end of this year, LEP1 will have ultimated its last run.
Although SLC will keep performing for a few more years, with some
(potentially, extremely interesting) longitudinal polarization
asymmetries \cite{Apol} still to be investigated, one can conclude that
the high precision SM test program, based on measurements of
electron-positron annihilation into fermions on top of Z resonance, has
essentially been concluded. No deviations from the SM prediction
was found (with the only possible
remarkable exception of the partial Z decay width into $b\bar b$)
\cite{LEP} \cite{Zbb} \cite{AGCbb}, at the
achieved precision level of few permille,
and for a large set of
experimental variables (partial and total Z widths and asymmetries)
that have been extremely carefully analyzed in these years. Stated
otherwise, and keeping in mind the previous remark, no virtual effects
of new physics have been evidentiated by the several measurements at
the few permille level performed in the considered four-fermion process
at $\sqrt{q^2} =M_Z$ ($q^2=(P_e+P'_e)^2$). Thus, the only information
achieved on several candidate models comes from a number of bounds,
that can be, depending on the case, drastic (e.g for most common
technicolour models \cite{TC}) or extremely mild (e.g. for the simplest
supersymmetric extension -MSSM\cite{MSSM}- of the SM).\par
Technicolour and supersymmetry are not the only alternatives to the SM
whose virtual effects have been tested by LEP1 and SLC. In particular,
signals of models with "anomalous" gauge couplings\cite{AGC} have been
also searched for and the negative results have led to some
corresponding bounds. But in this case the obtained results are somehow
less clean \cite{Hag-z}. Leaving aside a number of technical points, one
difficulty for these models is also related to the number
of involved parameters, essentially too large, even in the presence of
the considerable number of LEP1, SLC high precision measurements.\par
In a near future, electron-positron annihilation at
$\sqrt{q^2}\simeq 2M_Z$
(LEP2) and (perhaps in a "not too near" future) at $\sqrt{q^2}=500 GeV$
(NLC) will be measured. For obvious reasons, the relative accuracy of
the various measurements will be worse than at LEP1, SLC, moving from
the few permille to the few percent level. For final
fermion-antifermion state it is also likely that the number of
measurable experimental variables will decrease. This might lead to the
conclusion that the search of virtual effects of new physics in these
future processes will be, least to say, tough for a number of
potentially interesting models, in particular for those whose effect
\underline{on
top of Z resonance} are described by a large number of
parameters.\par
 In a previous paper \cite{Zsub}, we have tried to propose a solution
to this problem for the case of final (charged) lepton-antilepton
states. Our starting point was the (known) fact that a theoretical
analysis of virtual one loop effects can be eased by a proper choice of
the "input" parameters. For instance, for the description of physics of
electron-positron \underline{on Z resonance}, the introduction of the
Fermi constant $G_{\mu}$ to replace $M_W$ is quite useful. But for a
theoretical description of electron-positron annihilation at higher
energies, we showed in \cite{Zsub} that $G_{\mu}$ does not seem to be
the best choice if an investigation of models of new physics is the
theoretical goal. In particular, a self-contained representation of
final lepton-antilepton states can be given where $G_{\mu}$ is "traded"
and the new input parameters are the $Z$ leptonic width $\Gamma_l$ and
the "effective" $s^2_{eff}(M^2_Z)$ measured by LEP1 and SLC. Once these
quantities (together with the physical electric charge
$\alpha_{QED}(0)$) are introduced, the rest of the representation only
contains three subtracted quantities (called $\tilde{\Delta}\alpha$,
$R$ and $V$ in ref.\cite{Zsub}) whose theoretical properties for what
concerns the effects of a number of models of new physics appear
undeniably, least to say, interesting, in the sense that their actual
calculation turns out to be generally much easier (this is simply due
to the fact that a number of model's parameters, whose theoretical
features might be less pleasant, are often "reabsorbed" in the new
$Z$-peak inputs $\Gamma_l$ and $s^2_{eff}(M^2_Z)$). We also showed in
ref.\cite{Zsub} that, at the realistic expected experimental conditions
of future $e^+e^-$ colliders, the loss of theoretical accuracy
introduced by this "$G_{\mu}$ trading" does not produce any observable
effect. We concluded that the proposed "$Z$-peak subtracted"
representation was a good approach to investigate new-physics effects
in the final lepton-antilepton channel.\par
The aim of this paper is that of showing that the same method, with
identical conclusions, can be generalized to the case of final
quark-antiquark states. The new input parameters will be now those
hadronic quantities that are measured on $Z$ resonance (hadronic widths
and asymmetries, plus the strong coupling $\alpha_s(M_Z)$). Again, the
use of these inputs will allow to express the remaining one loop
theoretical expression in terms of \underline{subtracted} quantities,
that will be the three \underline{universal} corrections
$\tilde{\Delta}\alpha$,
$R$ and $V$ already met for the leptonic case and new, non universal
terms whose theoretical expression will be given for a number of
potentially interesting experimental quantities. This will be done in
full detail in the nextcoming Section 2. In the following Sections, we
shall try to show that the same remarkable features exhibited by our
representation for final leptonic states survive when one moves to
hadronic states. With this aim, we shall consider in Section 3
an example of "universal" effects of a model with
anomalous gauge couplings, showing that our method would help to solve
the problem of "parameters excess" for this case. We shall
also illustrate
how the possible experimental visible signatures of this model
would differ from those of another universal model of Technicolour
type. In Section 4, we
shall consider the example of "non universal" effects of a model with
one extra Z of the most general nature, and show that it can be
formally treated
as a special case of our approach. We shall
compare the effects of these models on a number of
observables and show that it is possible to select a special set of
three measurements that, in case a certain "signal" were observed,
would be able to indicate to which of the models it did
belong. In Section
5 a final discussion will show that for all the new input parameters,
the already available LEP1, SLC accuracy is sufficient to avoid the
generation of sensible uncertainties in the theoretical predictions at
the expected future experimental conditions. This will then conclude
our work.

\section{Derivation of the theoretical expressions}

{\bf a) General case}

\vspace{0.3cm}

The relevant quantity in our description will be the scattering
amplitude for the four-fermion process $e^+e^- \to f\bar f$ at variable
c.m. energy $\sqrt{q^2}$. A very convenient way of writing it at the
considered one loop level has been shown in ref.\cite{Zsub} for the
simplified case $f=l$. In this more general paper we shall begin
therefore by rewriting the needed expression, that reads:

 \bq \A^{(1)}_{lf}(q^2, \theta)= \A^{\gamma(1)}_{lf}(q^2, \theta)
+ \A^{Z(1)}_{lf}(q^2, \theta) \ \ + \ "QED" + \  "QCD" \eq

\noindent
with the photon exchange term

 \bq \A^{\gamma(1)}_{lf}(q^2, \theta) = \A^{\gamma(0)}_{lf}(q^2, \theta)
[1-\tilde{
F}^{(lf)}_{\gamma}(q^2, \theta)] \eq

 \bq \A^{\gamma(0)}_{lf}(q^2, \theta) \equiv {i\over q^2}
v_{\mu l}^{(\gamma)} \, v^{(\gamma) \mu f} = {ie^2_0Q_lQ_f\over q^2}
\bar v_l\gamma_{\mu} u_l \bar u_f\gamma^{\mu} v_f \eq

\noindent
($Q_{l,f}$ being the fermion charges in units of $|e|$)\par

\noindent
and the $Z$ exchange term

\bq  \A^{Z(1)}_{lf}(q^2, \theta) = {i\over
q^2-M^2_{Z0}}({g^2_0\over4c^2_0})[1- {A^{(lf)}_{Z}(q^2, \theta)\over
q^2-M^2_{Z0}}]\bar v_l\gamma^{\mu}(g^{(1)}_{Vl}-\gamma^5
g^{(0)}_{Al})u_l
 \bar u_f\gamma_{\mu}(g^{(1)}_{Vf}-\gamma^5 g^{(0)}_{Af})v_f\eq

\noindent
where

 \bq g^{(1)}_{Vl} = g^{(0)}_{Vl} -2s_1c_1Q_l
\tilde{F}^{(lf)}_{\gamma Z}(q^2, \theta) \eq

 \bq g^{(1)}_{Vf} = g^{(0)}_{Vf} -2s_1c_1Q_f
\tilde{F}^{(lf)}_{Z\gamma}(q^2, \theta) \eq

\noindent
with $s^2_1 = 1 - c^2_1$ and $s^{2}_{1} c^{2}_{1}
= \frac{\pi \alpha}{\sqrt{2}
G_{\mu}M^{2}_{Z}}$

Note that in the above equations bare couplings
$g_0=e_0/s_0$ ; $g^{(0)}_{Al,f} \equiv I^{3L}_{l,f}$ ;
$g^{(0)}_{Vl,f}\equiv  I^{3L}_{l,f} -2Q_{l,f}s^{2}_{0}$
and the bare Z mass $M_{Z0}$ are still
contained.\par
 The definition of the "generalized" one loop corrections,
that are \underline{gauge invariant} combinations of self-energies,
vertices and boxes belonging to the independent Lorentz structures of
the process (for a full and rigorous discussion about the choice of
gauge-invariant combinations, we defer to previous papers by Degrassi
and Sirlin \cite{Degrass}, to whose conclusions and notations we shall
try to stick as much as possible here) is the following:

\bq \tilde{F}^{lf}_{\gamma} (q^{2}, \theta) = F_{\gamma} (q^{2}) -
(\Gamma_{\mu,l}^{(\gamma)} , v_{\mu,l}^{(\gamma)}) -
(\Gamma_{\mu,f}^{(\gamma)} , v_{\mu,f}^{(\gamma)})+
A^{(B)}_ {\gamma, lf} (q^{2}, \theta) \eq

\bq A^{(lf)}_{Z} (q^{2}, \theta) = A_Z (q^{2}) - (q^2-M^2_{Z})[
(\Gamma_{\mu,l}^{(Z)} , v_{\mu,l}^{(Z)}) +
(\Gamma_{\mu,f}^{(Z)} , v_{\mu,f}^{(Z)})+
A^{(B)}_ {Z, lf} (q^{2}, \theta)] \eq

\bqa && \tilde{F}^{lf}_{\gamma Z} (q^{2}, \theta) \equiv
 {A^{(lf)}_{\gamma Z} (q^{2}, \theta)\over q^2}\nonumber\\
&& = {A_{\gamma Z} (q^{2})\over q^2} - {q^2-M^2_{Z}\over q^2}
(\Gamma_{\mu,f}^{(\gamma)} , v_{\mu,f}^{(Z)}) -
(\Gamma_{\mu,l}^{(Z)} , v_{\mu,l}^{(\gamma)})-(q^2-M^2_{Z})
A^{(B)}_ {\gamma Z, lf} (q^{2}, \theta) \eqa

\bqa && \tilde{F}^{lf}_{Z\gamma} (q^{2}, \theta) \equiv
 {A^{(lf)}_{Z\gamma} (q^{2}, \theta)\over q^2}\nonumber\\
&& = {A_{\gamma Z} (q^{2})\over q^2} - {q^2-M^2_{Z}\over q^2}
(\Gamma_{\mu,l}^{(\gamma)} , v_{\mu,l}^{(Z)}) -
(\Gamma_{\mu,f}^{(Z)} , v_{\mu,f}^{(\gamma)})-(q^2-M^2_{Z})
A^{(B)}_ {Z\gamma, lf} (q^{2}, \theta) \eqa

Here $F_{\gamma}$, $A_Z$, $A_{\gamma Z}$ are the conventional
self-energies, for which we shall follow the usual definition:

\bq A_{i}(q^{2}) \equiv A_{i}(0) + q^{2} F_{i} (q^{2}) \eq

\noindent
(note that the physical Z mass $M_Z$ now appears in eqs.(7)-(10)).
The quantities denoted as
$(\Gamma_{\mu l,f}^{(\gamma, Z)} , v_{\mu l,f}^{(\gamma, Z)})$
are the "components" of the generalized
$l$ (or $f$) vertex along the photon, or Z, Lorentz structure. For
instance, we would write for the overall photon vertex correction to a
fermion "f":

\bq \Gamma_{\mu f}^{(\gamma)} \equiv
(\Gamma_{\mu f}^{(\gamma)} , v_{\mu f}^{(\gamma)}) \,
v_{\mu f}^{(\gamma)} + (\Gamma_{\mu f}^{(\gamma)} , v_{\mu f}^{(Z)})
v_{\mu f}^{Z}  \eq

\noindent
where

 \bq v_{\mu f}^{(\gamma)}=
e_0Q_f\bar u_f\gamma_{\mu}v_f \eq

 \bq v_{\mu f}^{(Z)}=
{e_0\over 2c_0s_0}
\bar u_f\gamma_{\mu}(g^{(0)}_{Vf}-\gamma^5 g^{(0)}_{Af})v_f \eq

In our approach we shall need, rather than the previously defined
"generalized" corrections, the four "subtracted" quantities defined as:

\bq \tilde{\Delta}^{(lf)}\alpha(q^2,\theta) \equiv
\tilde{F}^{(lf)}_{\gamma} (0, \theta)
- \tilde{F}^{(lf)}_{\gamma} (q^{2}, \theta)  \eq

\bq R^{(lf)}(q^2,\theta) \equiv \tilde{I}^{(lf)}_Z (q^{2}, \theta) -
\tilde{I}^{(lf)}_Z (M^2_Z, \theta) \eq

\bq V^{(lf)}_{\gamma Z}(q^2,\theta) \equiv \tilde{F}^{(lf)}
_{\gamma Z} (q^{2}, \theta) -
\tilde{F}^{(lf)}_{\gamma Z} (M^2_Z, \theta) \eq

\bq V^{(lf)}_{Z\gamma}(q^2,\theta) \equiv \tilde{F}^{(lf)}
_{Z\gamma} (q^{2}, \theta) -
\tilde{F}^{(lf)}_{Z\gamma} (M^2_Z, \theta) \eq

\noindent
where the "auxiliary" quantity $I_Z$ is defined as

\bq \tilde{I}^{(lf)}_Z(q^2,\theta) =
{q^2\over q^2-M^2_Z}[\tilde{F}^{(lf)}_Z (q^2, \theta)-
\tilde{F}^{(lf)}_{Z} (M^2_Z, \theta)]  \eq

In eq.(1) the "pure QED" and the "pure QCD" components can be tested
separately and will not affect our research, which is only devoted to
the investigation on new \underline{electroweak} physics effects to the
one-loop perturbative order.\par

After these (we hope not too long) introductory definitions, that we
have given to make this paper as self-contained as possible, we are now
in a position to derive our general expressions.\par
The simplest way to illustrate the philosophy of our procedure is that
of showing the standard final form of the pure photonic contribution to
the scattering amplitude. Using the conventional definition of the
physical electric charge $\alpha\equiv \alpha(0)$ one immediately
realizes that:

\bq  {e^2_0\over q^2}[1-\tilde{F}^{(lf)}_{\gamma} (q^{2}, \theta)]
\equiv
{4\pi\alpha\over q^2}[1 + \tilde{\Delta}^{(lf)}\alpha(q^2,\theta)] \eq

\noindent
showing the known fact that the replacement of the bare charge by the
physical charge, \underline{measured at $q^2=0$}, is accompanied by the
replacement of the "generalized" correction $\tilde{
F}^{(lf)}_{\gamma}(q^2)$ with the "photon-peak" subtracted quantity
 $\tilde{ F}^{(lf)}_{\gamma}(q^2)-\tilde{ F}^{(lf)}_{\gamma}(0)$.\par
Our approach is based on a quite similar attitude for what concerns the
"pure $Z$" and the "$Z-\gamma$ interference" contributions. A typical
quantity to be considered corresponds e.g. in our conventions to the
term :

\bq  {ig^2_0\over4c^2_0}({1\over q^2-M^2_{Z0}})
[1- {A^{(lf)}_{Z}(q^2, \theta)\over
q^2-M^2_{Z0}}]  \eq

Using the tree level identity

\bq {g^2_0\over4c^2_0} = \sqrt2 G_{\mu,0} M^2_{Z,0} \eq

\noindent
one immediately realizes that the term eq.(21) becomes exactly:

\bqa && {g^2_0\over4c^2_0}({1\over q^2-M^2_{Z0}})
[1- {A^{(lf)}_{Z}(q^2, \theta)\over
q^2-M^2_{Z0}}] \equiv \nonumber\\
&& {\sqrt2 G_{\mu} M^2_Z\over
q^2-M^2_Z+iM_Z\Gamma_Z}[1+{\delta G_{\mu}\over G_{\mu}} +
Re {A^{(lf)}_{Z}(0, \theta)\over M^2_Z} -
\tilde{I}^{(lf)}_Z(q^2,\theta)] \eqa

\noindent
(with the conventional definition of $\Gamma_Z$).\par
In eq.(23) the physical input is represented by $M_Z$ and $G_{\mu}$,
with a certain "generalized" correction. Our approach consists,
essentially, in rewriting this term (and other, similar, ones) by
adding and subtracting $\tilde {I}_Z(M^2_Z, \theta)$. In the specific
case of eq.(23), this generates the quantity

\bqa && G_{\mu}[1+{\delta G_{\mu}\over G_{\mu}} +
Re {A^{lf}_{Z}(0, \theta)\over M^2_Z} -
\tilde{I}^{(lf)}_Z(M^2_Z,\theta)-\tilde{I}^{(lf)}_Z(q^2,\theta)
+\tilde{I}^{(lf)}_Z(M^2_Z,\theta)]\nonumber\\
&&=G_{\mu}[1+\epsilon^{(lf)}_1][1-
R^{(lf)}(q^2,\theta)] \eqa

\noindent
where $R^{(lf)}(q^2,\theta)$ is the
"Z-peak" subtracted correction defined by eq.(16) and

\bq \epsilon^{(lf)}_1=\epsilon_1 +
[(\Gamma_{\mu,f}^{(Z)} , v_{\mu,f}^{(Z)})-
(\Gamma_{\mu,l}^{(Z)} , v_{\mu,l}^{(Z)})]\eq

Here $\epsilon_1$ is the Altarelli-Barbieri parameter \cite{AB},
directly related to the partial Z width into leptons

\bq \Gamma_l = ({\sqrt2 G_{\mu}
M^3_{Z}\over48\pi})[1+\epsilon_1][1+\tilde{v}^2_l(M^2_Z)]\eq

\noindent
where

\bq \tilde{v}_l  \equiv 1-4s^2_l(M^2_Z) \eq

\noindent
and $s^2_l(M^2_Z)$ is the quantity measured at LEP1 and SLC.
Although a few more steps are still required, one can already
understand the final goal of our approach, i.e. that of "trading"
$G_{\mu}$ for some physical $Z$ partial width, measured at Z-peak. At
the same time, this procedure will replace the 'generalized"
corrections $\tilde{ I}^{(lf)}_Z(q^2)$ (and, also,
$\tilde{F}^{(lf)}_{\gamma Z}(q^2,\theta)$) with
the "Z-peak subtracted" quantities $R^{(lf)}(q^2,\theta)$,
$V^{(lf)}_{\gamma Z}(q^2,\theta)$,
$V^{(lf)}_{Z\gamma}(q^2,\theta)$ defined by
eqs.(16)-(18).\par

To fully understand the "replacement" mechanism, we shall now write the
complete one-loop expression of a representative observable,
chosen to be
$\sigma_{lf}(q^2)$, the cross section for production of a final $f\bar
f$ state. This can be done rather easily if one writes the tree-level
expression of this quantity, that reads:

\bqa && \sigma^{(0)}_{lf}(q^2)=N_f({4\pi q^2\over3})\{
{\alpha^2_0 Q^2_l Q^2_f\over q^4}+ \nonumber\\
&& [{\sqrt2 G^{(0)}_{\mu}
M^2_{Z0}\over4\pi}]^2{(g^{(0)\, 2}_{Vl}+g^{(0\, 2}_{Al})
(g^{(0)\, 2}_{Vf}+g^{(0)\, 2}_{Af})\over(q^2-M^2_{Z0})^2}+ \nonumber\\
&& 2{\alpha_0 Q_l
Q_f\over q^2}[{\sqrt2 G^{(0)}_{\mu}
M^2_{Z0}\over4\pi}]
{g^{(0)}_{Vl}g^{(0)}_{Vf}\over q^2-M^2_{Z0}}\} \eqa

\noindent
($N_f=3$ for quarks).

The corresponding expression at one loop is written immediately if one
uses our starting eq.(1) and makes the simple and obvious replacements.
One then easily derives:

\bq \sigma^{(1)}_{lf}(q^2)= \sigma^{(\gamma)}_{lf}(q^2)+
\sigma^{(Z)}_{lf}(q^2)+
\sigma^{(\gamma Z)}_{lf}(q^2) \ \ + \ "QED" + \  "QCD" \eq

\noindent
and finds for the "pure" Z exchange term:

\bqa && \sigma^{(Z)}_{lf}(q^2)= N_f({4\pi q^2\over3})
[{\sqrt2 G_{\mu}
M^2_{Z}\over16\pi}]^2({1\over(q^2-M^2_Z)^2+M^2_Z\Gamma^2_Z})[1+
2\epsilon^{(lf)}_1]
[1+\tilde{v}^2_l]
[1+\tilde{v}^2_f]\nonumber\\
&& [1-2\tilde{I}^{(lf)}_Z(q^2,\theta)
 -8s_1c_1\{{v_1\over1+v^2_1}V^{(lf)}_{\gamma
Z}(q^2)+{v_f |Q_f|\over1+v^2_f}V^{(lf)}_{Z\gamma}(q^2)\}] \eqa

\noindent
where $v_1$, $v_f$ are defined as

\bq v_{1,f}=1-4|Q_{l,f}|s^2_1 \eq

\noindent
and $s^2_1=0.212$ is defined after eqs.(2),(3).\par
We also define the quantity (not to be confused with the one above )

\bq \tilde{v}_{f}=1-4|Q_{f}|s^2_{f}(M^2_Z) \eq

\noindent
with

\bq s^2_f(M^2_Z)\equiv s^2_1 + s_1c_1\tilde{F}^{(lf)}_{Z\gamma}(M^2_Z)
\eq

{}From eq.(30) one recovers the result of ref.\cite{Zsub}
when $f=l$. In that
case, $v_1=v_f$, $\epsilon_{1,lf}=\epsilon_1$, $F_{\gamma
Z}=F_{Z\gamma}$ and in the "leading" terms $(G_{\mu})^2$ has been
exactly replaced by the quantity $[{\Gamma_l\over M_Z}]^2$ (multiplied
by a c-number), while in the correction the "Z-peak subtracted"
quantities $R\equiv R^{ll}$ and $V\equiv V^{ll}_{\gamma Z}\equiv
V^{ll}_{Z\gamma}$ appear.\par
When $l\neq f$, a very similar situation can be reproduced. One only
has to introduce the quantity:

\bq \epsilon^{ff}_1 = \epsilon_1 + 2
[(\Gamma_{\mu,f}^{(Z)} , v_{\mu,f}^{(Z)}) -
(\Gamma_{\mu,l}^{(Z)} , v_{\mu,l}^{(Z)})]  \eq

This is \underline{exactly} related to the partial width of $Z$ into
$f\bar f$ by the following expression \cite{vertex}

\bq \Gamma_f = N^{QCD}_f({\sqrt2 G_{\mu}
M^3_{Z}\over48\pi})[1+\epsilon^{ff}_1][1+\tilde{v}^2_f(M^2_Z)] \eq

\noindent
where

\bq N^{QCD}_f \simeq 1 + {\alpha_s(M^2_Z)\over\pi} \eq

The final observation is now the exact equality:
 \bq 2\epsilon^{(lf)}_1\equiv \epsilon_1 + \epsilon^{(ff)}_1   \eq

{}From this equality and from the previous formulae one is then finally
led to the relevant expression:

\bqa && \sigma^{(Z)}_{lf}(q^2)=N_f({4\pi q^2\over3}){[{3\Gamma_l\over
M_Z}][{3\Gamma_f\over N_f M_Z}]\over(q^2-M^2_Z)^2+M^2_Z\Gamma^2_Z}[
1-2R^{(lf)}(q^2)\nonumber\\
&& -8s_1c_1\{{v_1\over1+v^2_1}V^{(lf)}_{\gamma
Z}(q^2)+{v_f |Q_f|\over1+v^2_f}V^{(lf)}_{Z\gamma}(q^2)\}] \eqa

Eq.(38) is one of the main results of this paper. It shows that the
replacement of $G_{\mu}$, and the corresponding introduction of "Z-peak
subtracted" corrections, can be continued to final hadronic states by
introduction of quantities that correspond to those encountered in the
leptonic case. Typically, $\sigma_{lf}$ will contain $\Gamma_l$
\underline{and} $\Gamma_f$, as one would have naively expected, and the
strong coupling $\alpha_S(M^2_Z)$ generated by eq.(35), that only
affects the expression if $l\neq f$ and should \underline{not} be
considered as a "QCD" correction in the notation of eq.(1). Note
that only quantities that can be exactly defined and (in principle)
\underline{measured} on Z resonance have been used to build our "Z-peak
modified Born approximation".\par
The procedure that we have illustrated
can now be repeated for the remaining components of $\sigma_{lf}$ (as
well as for the other observables). In fact, there is no need of any
trick for the "pure $\gamma$" component, that remains given by the
expression:

\bq \sigma^{(\gamma)}_{lf}(q^2)=N_f({4\pi q^2\over3})Q^2_l Q^2_f
{\alpha^2(0)\over q^4}[1+2\tilde{\Delta}^{(lf)}\alpha(q^2)] \eq

The "$\gamma-Z$" interference can be treated in a straightforward way.
To avoid writing too many formulae, we only give here the relevant
final expression, that can be easily derived using the previously
illustrated procedure:

\bqa && \sigma^{(\gamma Z)}_{lf}(q^2)=N_f({4\pi q^2\over3})
2\alpha(0)|Q_f|{q^2-M^2_Z\over
q^2(q^2-M^2_Z)^2+M^2_Z\Gamma^2_Z)}[{3\Gamma_l\over
M_Z}]^{1/2}[{3\Gamma_f\over N_f M_Z}]^{1/2}\nonumber\\
 && {\tilde{v}_l \tilde{v}_f\over
(1+\tilde{v}^2_l)^{1/2}(1+\tilde{v}^2_f)^{1/2}}
[1+
\tilde{\Delta}^{(lf)}\alpha(q^2) -R^{(lf)}(q^2)\nonumber\\
 && -4s_1c_1
\{{1\over v_1}V^{(lf)}_{\gamma Z}(q^2)+{|Q_f|\over v_f}
V^{(lf)}_{Z\gamma}(q^2)\}] \eqa

Note that, besides $\Gamma_l$ and
$\Gamma_f$, this expression contains the two
parameters $s^2_l(M^2_Z)$ and $s^2_f(M^2_Z)$ (or $\tilde{v}_l$ and
$\tilde{v}_f$) defined in eqs.(27), (32), (33),
that are now \underline{not}  reabsorbed into $\Gamma_l$,
$\Gamma_f$ as in the previous "pure Z" term. This does not represent a
problem since $s^2_l(M^2_Z)$ is the quantity measured at LEP1, SLC. The
remaining parameter $s^2_f(M^2_Z)$ is also related to measured (or
measurable) quantities at Z-peak, more precisely to forward-backward
unpolarized asymmetries for $b$ and $c$ (already given by LEP1) and
also, more directly, to the polarized forward-backward asymmetries for
$b$ and $c$, called $A_{b,c}$ in the original proposal \cite{Apol},
to be
measured at SLC in the near future \cite{ASLC}. In particular, in terms
of $A_{b,c}$ we would have

\bq A_{b,c} = {2\tilde{v}_{b,c}\over 1 + \tilde{v}^2_{b,c}}\eq

\noindent
(the unpolarized asymmetries are essentially given by the product of
eq.(41) with the corresponding leptonic quantity that contains
$s^2_l(M^2_Z)$). In conclusion, also in the case of $\sigma^{\gamma
Z}$, the new complete "Born" expression can be given in terms of
quantities measured on $Z$ resonance. As we shall show in the final
discussion, this will never introduce a relevant "input" uncertainty in
the obtained predictions.\par
To conclude this general part of Section 2, we still need the
derivation of the quantity that appears in the numerator of an
\underline{unpolarized} forward-backward asymmetry. We shall write this
observable in the following way:

\bq A_{FB,f}(q^2) = {3\sigma_{FB,lf}(q^2)\over 4\sigma_{lf}(q^2)}  \eq

\noindent
where $\sigma_{lf}$ has been previously defined. From the expression
(that we do not write explicitely) of $\sigma_{FB,lf}$ at tree level it
is immediate to derive, without introducing any other prescription or
definition, the final relevant expression:

\bq \sigma_{FB,lf}(q^2) \equiv \sigma^{(Z)}_{FB,lf}(q^2) +
\sigma^{(\gamma Z)}_{FB,lf}(q^2) \eq

\noindent
where:

\bqa && \sigma^{(Z)}_{FB,lf}(q^2)=N_f({4\pi q^2\over3}){[{3\Gamma_l\over
M_Z}][{3\Gamma_f\over N_f
M_Z}]\over(q^2-M^2_Z)^2+M^2_Z\Gamma^2_Z}\nonumber\\
&& {4\tilde{v}_l \tilde{v}_f\over(1+\tilde{v}^2_l)(1+\tilde{v}^2_f)}[
1-2R^{(lf)}(q^2)-4s_1c_1
\{{1\over v_1}V^{(lf)}_{\gamma Z}(q^2)+{|Q_f|\over v_f}
V^{(lf)}_{Z\gamma}(q^2)\}]  \eqa

\noindent
and

\bqa && \sigma^{(\gamma Z)}_{FB,lf}(q^2)=N_f({4\pi q^2\over3})
2\alpha(0)|Q_f|{q^2-M^2_Z\over
q^2((q^2-M^2_Z)^2+M^2_Z\Gamma^2_Z)}\nonumber\\
&& [{3\Gamma_l\over
M_Z}]^{1/2}[{3\Gamma_f\over N_f
M_Z}]^{1/2}{1\over(1+\tilde{v}^2_l)^{1/2}(1+\tilde{v}^2_f)^{1/2}}[1+
\tilde{\Delta}^{(lf)}\alpha(q^2) -R^{(lf)}(q^2)] \eqa

Eqs.(44) and (45) conclude our general technical introduction. We
shall now consider in the next subsection 2b the explicit cases of
experimental observables that will, or should, be measured in the very
near future at LEP2 and, possibly, in a not too near future at NLC.\par

{\bf  2b)Application to specific observables}

\vspace{0.3cm}

To begin our analysis, we consider the simplest case that might
realistically occur, i.e. that of the measurement of the cross section
for $b\bar b$ production, $\sigma_{lb}$. Actually, we should rather
consider the (experimentally more accurate) ratio
$R_{bl}=\sigma_{lb} / \sigma_{ll}$. Since the theoretical expression of
$\sigma_{ll}$ has been already given in ref.\cite{Zsub}, we shall limit
ourselves to deriving and discussing in detail the full expression of
the numerator. Then in Sections 3 and 4 we shall rather use the
ratio, whose expression can be easily derived.\par
When writing the full expression of $\sigma_{lb}$, as well as that of
the next considered observables, it will be very useful to separate the
"universal" contributions of new physics from the "non universal" ones,
that depend on properties of the final state that are different from
the corresponding ones for leptons (e.g. specific non SM couplings, or
masses). Clearly, the full set of self-energies contributions will
belong to the first universal class, while boxes will generally produce
non universal effects. For vertices, one can have both cases, as we
shall show in the next sections.\par
After these premises, we can now write the complete expression

\bq \sigma_{lb}(q^2) = \sigma^{(\gamma)}_{lb}(q^2) +
\sigma^{(Z)}_{lb}(q^2)
+ \sigma^{(\gamma Z)}_{lb}(q^2)\eq

\noindent
where the three components are given by eqs.(39),(44),(45) and,
following our previous discussion, we shall express the subtracted
corrections in the form:

\bq \tilde{\Delta}^{(lb)}\alpha(q^2) =
\tilde{\Delta}\alpha(q^2) +
\delta\tilde{\Delta}^{(lb)}\alpha(q^2) \eq

\bq R^{(lb)}(q^2) = R(q^2) + \delta R^{(lb)}(q^2) \eq

\bq V^{(lb)}_{\gamma Z}(q^2) = V(q^2) +
\delta V^{(lb)}_{\gamma Z}(q^2) \eq

\bq V^{(lb)}_{Z\gamma}(q^2) = V(q^2) +
\delta V^{(lb)}_{Z\gamma}(q^2) \eq

\noindent
where the quantities without indices are the universal ones that would
appear in the case of final leptonic states treated in \cite{Zsub} and :

\bqa && \delta\tilde{\Delta}^{(lb)}\alpha(q^2) =
(\Gamma_{\mu,l}^{(\gamma)}(0) , v_{\mu,l}^{(\gamma)})-
(\Gamma_{\mu,b}^{(\gamma)}(0) , v_{\mu,b}^{(\gamma)})-
[(\Gamma_{\mu,l}^{(\gamma)}(M^2_Z) , v_{\mu,l}^{(\gamma)})-
(\Gamma_{\mu,b}^{(\gamma)}(M^2_Z) , v_{\mu,b}^{(\gamma)})]\nonumber\\
&& +
A^{(B)}_ {\gamma, ll} (q^{2}, \theta)-
A^{(B)}_ {\gamma, lb} (q^{2}, \theta)\eqa

\bqa &&  \delta R^{(lb)}(q^2) =
Re\{(\Gamma_{\mu,l}^{(Z)}(q^2) , v_{\mu,l}^{(Z)})-
(\Gamma_{\mu,b}^{(Z)}(q^2) , v_{\mu,b}^{(Z)})-
[(\Gamma_{\mu,l}^{(Z)}(M^2_Z) , v_{\mu,l}^{(Z)})-
(\Gamma_{\mu,b}^{(Z)}(M^2_Z) , v_{\mu,b}^{(Z)})]\nonumber\\
&& +
A^{(B)}_ {Z, ll} (q^{2}, \theta)-
A^{(B)}_ {Z, lb} (q^{2}, \theta)
\}\eqa

\bqa && \delta V^{(lb)}_{\gamma Z}(q^2) =
{q^2-M^2_Z\over q^2}Re[(\Gamma_{\mu,b}^{(\gamma)}(q^2)
, v_{\mu,b}^{(Z)})-
(\Gamma_{\mu,l}^{(\gamma)}(q^2) , v_{\mu,l}^{(Z)})]\nonumber\\
&& +
(q^2-M^2_Z)Re[A^{(B)}_ {\gamma Z, ll} (q^{2}, \theta)
 - A^{(B)}_ {\gamma Z, lb} (q^{2}, \theta)]\eqa

\bqa&&  \delta V^{(lb)}_{Z\gamma}(q^2) =
Re\{(\Gamma_{\mu,l}^{(Z)}(q^2) , v_{\mu,l}^{(\gamma)})-
(\Gamma_{\mu,b}^{(Z)}(q^2) , v_{\mu,b}^{(\gamma)})-
[(\Gamma_{\mu,l}^{(Z)}(M^2_Z) , v_{\mu,l}^{(\gamma)})-
(\Gamma_{\mu,b}^{(Z)}(M^2_Z) , v_{\mu,b}^{(\gamma)})]\nonumber\\
&& +(q^2-M^2_Z)[A^{(B)}_ {Z\gamma, ll} (q^{2}, \theta)
- A^{(B)}_ {Z\gamma, lb} (q^{2}, \theta)]
\} \eqa

\noindent
Note that by definition $A^{(B)}_ {\gamma, lf} (0, \theta)$ and
$A^{(B)}_ {Z, lf} (M^2_Z, \theta)$ identically vanish.\par
The previous expressions and definitions can be easily generalized to
the case of the full final hadronic cross section, whose experimental
measurement will be statistically favoured. After some additions and
recombinations we are led to a first general expression that would
read:
 \bq \sigma_5(q^2) \equiv \sigma_{l had}(q^2) = \sigma^{(\gamma)}_5(q^2) +
\sigma^{(Z)}_5(q^2) + \sigma^{(\gamma Z)}_5(q^2) \eq

\noindent
where

\bq  \sigma^{(\gamma)}_5(q^2)=N
({4\pi q^2\over3})({11\alpha^2(0)\over9q^4})[1+\delta^{(\gamma)}_5]\eq

\bq \delta^{(\gamma)}_5 =
2\tilde{\Delta}^{(ll)}\alpha(q^2) + {16\over11}
\tilde{\Delta}^{(lu)}\alpha(q^2) + {4\over11}
\tilde{\Delta}^{(ld)}\alpha(q^2) + {2\over11}
\tilde{\Delta}^{(lb)}\alpha(q^2)\eq

\noindent
where following the attitude explained at the beginning of Section 2a,
we set $N=3$. For the next term $\sigma^{(Z)}_5$ we find, after a
number of elementary steps

\bq  \sigma^{(Z)}_5(q^2)=N
({4\pi q^2\over3}){[{3\Gamma_l\over
M_Z}][{3\Gamma_5\over N_f M_Z}]\over(q^2-M^2_Z)^2+M^2_Z\Gamma^2_Z}
[1+\delta^{(Z)}_5]\eq

\noindent
where $\Gamma_5=\Gamma_{had}$ and

\bqa && \delta^{(Z)}_5 = -2R(q^2)-4s_1c_1p_5V(q^2)\nonumber\\
&& -{2\Gamma_u\over\Gamma_5}[2\delta R^{(lu)}(q^2)
+{8s_1c_1v_1\over1+v^2_1}\delta V^{(lu)}_{\gamma Z}(q^2)
+{16s_1c_1v_u\over3(1+v^2_u)}\delta V^{(lu)}_{Z\gamma}(q^2)]\nonumber\\
&& -{2\Gamma_d\over\Gamma_5}[2\delta R^{(ld)}(q^2)
+{8s_1c_1v_1\over1+v^2_1}\delta V^{(ld)}_{\gamma Z}(q^2)
+{8s_1c_1v_d\over3(1+v^2_d)}\delta V^{(ld)}_{Z\gamma}(q^2)]\nonumber\\
&& -{\Gamma_b\over\Gamma_5}[2\delta R^{(lb)}(q^2)
+{8s_1c_1v_1\over1+v^2_1}\delta V^{(lb)}_{\gamma Z}(q^2)
+{8s_1c_1v_b\over3(1+v^2_b)}\delta V^{(lb)}_{Z\gamma}(q^2)]\eqa

\noindent
with

\bq  p_5= {v_1\over1+v^2_1}+{4\Gamma_u\over3\Gamma_5}
({v_u\over1+v^2_u})+
{2\Gamma_d\over3\Gamma_5}({v_d\over1+v^2_d}) +
{\Gamma_b\over3\Gamma_5}({v_b\over1+v^2_b}) \eq

{}From a glance at eq.(59), one might have the impression that both in
the "universal" and in the
"non universal" component of the corrections a number of unwanted
(i.e. not directly
measured on Z resonance) ratios $\Gamma_q / \Gamma_5$ appear. But this
is not a problem at the considered one-loop level since these terms are
already multiplied by order ($\alpha$). Therefore, they must be
consistently replaced by expressions that only involve the
quantity $s^2_1$
entering eq.(33) (note that $\epsilon^{ff}_1$ can be neglected for the
same reasons). As a consequence we can write in eq.(59)

\bq  {\Gamma_{u,c} \over \Gamma_5} = {1+v^2_u\over
2(1+v^2_u)+3(1+v^2_d)}\eq

\bq  {\Gamma_{d,s,b} \over \Gamma_5} = {1+v^2_d\over
2(1+v^2_u)+3(1+v^2_d)}\eq

The same considerations and simplifications strictly valid at one loop
can be repeated for the interference component. After some
straightforward rearrangements, this leads to the expression:

\bqa  && \sigma^{(\gamma Z)}_5(q^2)=N({4\pi q^2\over3})
{2\over3}\alpha(0){q^2-M^2_Z\over
q^2(q^2-M^2_Z)^2+M^2_Z\Gamma^2_Z)}\nonumber\\
 && [{3\Gamma_l\over
M_Z}]^{1/2}\Sigma_5{\tilde{v}_l\over(1+\tilde{v}^2_l)^{1/2}}[1+
\delta^{(\gamma Z)}_5] \eqa

\bqa && \delta^{(\gamma Z)}_5 = \tilde{\Delta}\alpha(q^2)
 -R-4s_1c_1p'_5V\nonumber\\
&& +{4\over\Sigma_5}({3N_u\Gamma_u\over
M_Z})^{1/2}{v_u\over(1+v^2_u)^{1/2}})
[\delta\tilde{\Delta}^{(lu)}\alpha(q^2)-\delta R^{(lu)}(q^2)
-{4s_1c_1\over v_1}\delta V^{(lu)}_{\gamma Z}(q^2)
-{8s_1c_1\over3v_u}\delta V^{(lu)}_{Z\gamma}(q^2)]\nonumber\\
&& +{2\over\Sigma_5}({3N_d\Gamma_d\over
M_Z})^{1/2}{v_d\over(1+v^2_d)^{1/2}})
[\delta\tilde{\Delta}^{(ld)}\alpha(q^2)-\delta R^{(ld)}(q^2)
-{4s_1c_1\over v_1}\delta V^{(ld)}_{\gamma Z}(q^2)
-{4s_1c_1\over3v_d}\delta V^{(ld)}_{Z\gamma}(q^2)]\nonumber\\
&& +{1\over\Sigma_5}({3N_b\Gamma_b\over
M_Z})^{1/2}{v_b\over(1+v^2_b)^{1/2}})
[\delta\tilde{\Delta}^{(lb)}\alpha(q^2)-\delta R^{(lb)}(q^2)
-{4s_1c_1\over v_1}\delta V^{(lb)}_{\gamma Z}(q^2)
-{4s_1c_1\over3v_b}\delta V^{(lb)}_{Z\gamma}(q^2)]\nonumber\\
 \eqa

\noindent
with

\bq  \Sigma_5= 4({3N_u\Gamma_u\over
M_Z})^{(1/2)}{\tilde{v}_u\over(1+\tilde{v}^2_u)^{1/2}}+
2({3N_d\Gamma_d\over M_Z})^{(1/2)}{\tilde{v}_d\over(1+
\tilde{v}^2_d)^{1/2}} +
({3N_b\Gamma_b\over M_Z})^{(1/2)}{\tilde{v}_b\over(1+
\tilde{v}^2_b)^{1/2}} \eq

\bq  p'_5=  {1\over
v_1}+{8\over3(1+v^2_u)^{1/2}\Sigma_5}({3N_u\Gamma_u\over
M_Z})^{1/2} + {2\over3(1+v^2_d)^{1/2}\Sigma_5}({3N_d\Gamma_d\over
M_Z})^{1/2} + {1\over3(1+v^2_b)^{1/2}\Sigma_5}({3N_b\Gamma_b\over
M_Z})^{1/2}
 \eq
\noindent
and $N_{u,d,b}=3$.\par
Similarly to the case of eq.(61),(62), the
quantities $\Gamma^{1/2}_f/\Sigma_5$
can be safely evaluated in terms of $s^2_1$ only. The quantity
$\Sigma_5$ eq.(65) requires a separate discussion, that will be
given in the concluding remarks, to show that it can be safely
neglected (or approximated).\par
To conclude our review, we still have to consider the separation of the
quantities eq.(44),(45) that give the numerator of the forward-backward
asymmetry for $b\bar b$ production. This can be done in the way that we
have illustrated, and leads to the expressions

\bqa && \sigma^{(\gamma Z)}_{FB,lb}(q^2)=N({4\pi q^2\over3})
2\alpha(0)|Q_b|{q^2-M^2_Z\over
q^2(q^2-M^2_Z)^2+M^2_Z\Gamma^2_Z)}\nonumber\\
&& [{3\Gamma_l\over
M_Z}]^{1/2} [{3\Gamma_b\over
N_b M_Z}]^{1/2}
({1\over(1+\tilde{v}^2_l)^{1/2}(1+\tilde{v}^2_b)^{1/2}})\nonumber\\
 && [1+
\tilde{\Delta}\alpha(q^2)- R(q^2) +
\delta\tilde{\Delta}^{(lb)}\alpha(q^2)-
\delta R^{(lb)}(q^2)]
\eqa

\bqa && \sigma^{(Z)}_{FB,lb}(q^2)=N({4\pi q^2\over3}){[{3\Gamma_l\over
M_Z}][{3\Gamma_f\over N_f M_Z}]\over[(q^2-M^2_Z)^2+M^2_Z\Gamma^2_Z]}
({4\tilde{v}_l \tilde{v}_b\over(1+\tilde{v}^2_l)(1+\tilde{v}^2_b)})
\nonumber\\
&& [1-2R(q^2)-4s_1c_1
[{1\over v_1}+{|Q_b|\over v_b}]V(q^2)
\nonumber\\
 && -2\delta R^{(lb)}(q^2)-4s_1c_1\{{1\over v_1}
\delta V^{(lb)}_{\gamma Z}(q^2)
+{|Q_b|\over v_b}
\delta V^{(lf)}_{Z\gamma}(q^2)\}]  \eqa

Eqs.(67),(68) conclude this long section. We are now in a position to
calculate, using the leptonic formulae or ref.\cite{Zsub}, the
contributions of new physics of both universal and non universal type
to the full set on experimental quantities that will be measured at
LEP2 and NLC (without the extra facility of longitudinal initial
electron polarization in the latter case). In particular, we shall
consider on top of the leptonic observables previously considered in
ref.\cite{Zsub}, i.e. $\sigma_{\mu}$, $A_{FB,\mu}$ and $A_{\tau}$ (the
final $\tau$ polarization), the \underline{ratios}:

\bq R_5 = {\sigma_5\over \sigma_{\mu}} \eq

\bq R_b = {\sigma_{lb}\over \sigma_{\mu}} \eq

\noindent
and $A_{FB,b}$. The relevant expressions can be derived from Section 2
and from ref.\cite{Zsub}. We shall give them explicitely in the next
Section 3 and 4 for two orthogonal situations of models with
universal and non universal type of effects.

\section{A Model with AGC (Anomalous Gauge Couplings)}

As a first example of application of our approach, we shall consider
the case of a model of new physics in which Anomalous Gauge Couplings
\cite{AGC} are generated by an effective lagrangian. Although the
discussion could be much more general, we shall first stick to the
dimension six, CP conserving Lagrangian proposed by
Hagiwara, S. Ishihara, R. Szalapski
and D. Zeppenfeld \cite{Hag-z}.
This contains, in principle, eleven parameters
of which nine would affect the $WWV$ couplings. In particular, the most
general four fermion process at the one loop level would be affected by
four "renormalized" parameters denoted in
ref.\cite{Hag-z} as $f^r_{DW}$,
$f^r_{DB}$, $f^r_{\phi 1}$ and $f^r_{BW}$ for $f\neq b$. If final
$b\bar b$ production is considered, one should, in principle, include
the order ($m^2_t$) contributions generated by $f_W$, $f_B$ that have
been recently shown to appear in the partial width of $Z$ into $b\bar
b$ \cite{AGCbb}.\par
The calculation of this type of effects has been already performed for
the purely leptonic case in ref.\cite{Zsub}.
The main feature that appears
is that only \underline{two} independent parameters i.e.
$f^r_{DW}$, $f^r_{DB}$ survive in the full
set of leptonic observables. This is due to the fact that in the
contribution of the model to the subtracted corrections $\tilde
{\Delta}\alpha$, $R$ and $V$ the terms proportional to $f_{\phi 1}$,
$f_{BW}$, that carry no sufficient powers of $q^2$, are fully
reabsorbed into the subtraction constant i.e. into the trading of
$G_{\mu}$ by $\Gamma_l$ and $s^2_l(M^2_Z)$. This leads to the
expressions:

\bq \tilde{\Delta}^{(AGC)}\alpha(q^2)= -q^2({2e^2\over\Lambda^2})
(f^r_{DW}+f^r_{DB}) \eq

\bq  R^{(AGC)}(q^2)= (q^2-M^2_Z)({2e^2\over s^2_1c^2_1\Lambda^2})
(f^r_{DW}c^4_1+f^r_{DB}s^4_1) \eq

\bq V^{(AGC)}(q^2) = (q^2-M^2_Z)({2e^2\over s_1c_1\Lambda^2})
(f^r_{DW}c^2_1+f^r_{DB}s^2_1)  \eq

We now perform the same calculation for $R_5$, $R_b$ and $A_{FB,b}$. In
principle, we might expect the appearence of the extra parameters
$f_W$, $f_B$ in the expression of the final $b$ contribution. In fact,
the rigorous expression for $R_5$ would read (neglecting numerically
irrelevant contributions):

\bqa && {\delta R^{(AGC)}_5 \over  R_5} \simeq
C_{\alpha}(q^2) \tilde{\Delta}^{(AGC)}\alpha(q^2) +\nonumber\\
&& C_R(q^2) R^{(AGC)}(q^2) + C_V(q^2) V^{(AGC)}(q^2) +
C_b(q^2) \delta R^{(AGC)}_{lb}(q^2)  \eqa

\noindent
and $C_{\alpha}$, $C_R$, $C_V(q^2)$ are certain kinematical functions
whose numerical value at the "reference" points $q^2=4M^2_Z$ (LEP2) and
$q^2=(500 GeV)^2$ are:

\bq C_{\alpha}(4M^2_Z)= -0.77  \ \ \ \ \ \
C_{\alpha}((500 GeV)^2)= -0.67 \eq

\bq C_R(4M^2_Z)= -0.77   \ \ \ \ \ \
 C_R((500 GeV)^2)= -0.67\eq

\bq C_V(4M^2_Z)= - 0.81  \ \ \ \ \ \
 C_V((500 GeV)^2)= -0.75
\eq

The last term in eq.(77) contains a kinematical coefficient $C_b$ such
that

\bq C_b(4M^2_Z) = -0.25  \ \ \ \ \ \
 C_b((500 GeV)^2) = -0.20\eq

\noindent
and a "non universal" contribution, typical of the final $Zb\bar b$
couplings. In terms of parameters of the model, one gets after a
straightforward calculation whose main points have been illustrated in
a previous reference \cite{AGCbb}:

\bq \delta R^{(AGC)}_{lb} = 2({q^2-M^2_Z\over M^2_Z})({\alpha
m^2_t\over64\pi s^2_1\Lambda^2})(f_W-f_B{s^2_1\over c^2_1})
Log({\Lambda^2\over
M^2_Z}) \eq

In fact, in ref.\cite{AGCbb} a bound for a
different combination of $f_W$,
$f_B$ was calculated, assuming that the still conceivable small
discrepancy between the experimental value of $\Gamma_{b\bar b}$ at
resonance and the SM prediction was originated by this type of new
physics. From that calculation one sees however that, even pushing the
bound to the extreme value, we would not affect the relative $R_5$
shift by more than a fraction of a percent, hardly visible at realistic
experimental conditions. For this reason, and keeping in mind that the
complete correction to $R_5$ contains in principle such non universal
terms, we have in fact neglected them in the nextcoming considerations.
This has the welcome consequence that another experimental variable can
be added to the previous leptonic set without increasing the overall
number of parameters to be fitted, or bounded. More precisely, we would
have now, at LEP2 energy:

\bq {\delta R^{(AGC)}_5\over R_5} =
1.87({M^2_Z\over \Lambda^2})f^r_{DW}
+ 0.68({M^2_Z\over \Lambda^2})f^r_{DB}\eq

The previous considerations can be exactly repeated for $R_b$. Leaving
aside a more general discussion, we would find in this case in the
configuration $q^2=4M^2_Z$:

\bq {\delta R^{(AGC)}_b\over R_b} =
-1.13[\tilde{\Delta}^{(AGC)}\alpha(q^2)+ R^{(AGC)}(q^2)]
-0.94 V^{(AGC)}(q^2)-1.53\delta R^{(AGC)}_{lb}(q^2) \eq

Again, the conceivable contribution from the non universal term would
be, at most, of a few (two-three) relative percent, that should be
realistically below the observability limits. Neglecting again this
contribution would lead us to the approximate expression :

\bq {\delta R^{(AGC)}_b\over R_b} =
-4.14({M^2_Z\over \Lambda^2})f^r_{DW}
-1.26({M^2_Z\over \Lambda^2})f^r_{DB}\eq

To conclude this illustration, we have calculated the contribution to
the forward-backward $b$-asymmetry. This quantity, unlike the two
previous cases, does not receive in practice appreciable contributions
from the non universal part, which is essentially of left-handed type.
The rigorous expression at LEP2 energies would therefore read:

\bq {\delta A^{(AGC)}_{FB,b}\over A_{FB,b}} =
0.40[\tilde{\Delta}^{(AGC)}\alpha(q^2)+ R^{(AGC)}(q^2)]
-0.42 V^{(AGC)}(q^2)\eq

In conclusion, we have now at our disposal six experimental variables
($\sigma_{\mu}$, $A_{FB,\mu}$, $A_{\tau}$, $R_5$, $R_b$ and $A_{FB,b}$)
that only depend on \underline{two} parameters (and that, at most,
would contain \underline{one} extra third combination of $f_W$ and
$f_B$). This represents, in our opinion, an interesting alternative
to the conventional analyses
\cite{Hag-z}, where
the full set of six parameters should enter in the previous
observables. In fact a rigorous calculation,
that fully takes into account
the effects of QED radiation, is at the moment being performed and will
be shown in a separate dedicated paper. Here we can give a qualitative
hint looking e.g. at the particular effect on $R_5$, eq.(64). In
correspondence to a typical couple of values that would still be
allowed \cite{Hagnew} by the available low-energy constraints i.e.
$f_{DW}=-1$, $f_{DB}=4$, we would find a relative positive shift of
approximately six percent in $R_5$, that would lead to a spectacular
visible signal.\par
 As a final byproduct of our approach, in which the
number of parameters for this specific model is drastically reduced and
in practice only two independent quantities remain, we shall obtain
the (pleasant) result that, for any chosen triplet of observables,
there will be a linear relationship between the separate effects that
will correspond to a plane in the 3-dimensional
space of the observables.
Drawing these planes for various choices of variables is rather easy.
Here we want to show two particular examples related to the choices of
($\sigma_{\mu}$, $A_{FB,\mu}$, $A_{\tau}$) and
($\sigma_{\mu}$, $A_{FB,\mu}$, $R_5$) as "coordinate axes". The
corresponding regions are shown in Figs.1,2 in the simple
approximation that corresponds to our approximate equations (a more
rigorous derivation, with a full QED convolution of effects, will be
given, as we preannounced, in a forthcoming paper). To make a
meaningful statement, we have shown in these Figures the "dead" region
where a signal would not be distinguishable, corresponding to a
relative experimental error of 1.5 percent for the various cross
sections and forward-backward asymmetry and 15 percent for the tau
polarization (these values assume an integrated luminosity of 500
$pb^{-1}$ at $\sqrt{q^2} = 2M_Z$, and correspond to a muon cross
section of 4.4 $pb$). Therefore,
\underline{if} a signal of new physics were seen in some of the
aforementioned observables, one would be able to decide whether the
signal belongs to the considered model, or not. In fact, one might even
hope to find a sort of one-to-one correspondence between models and
regions of a certain 3-dimension space of observables.\par
Although we
cannot prove this statement in general, we have found an encouraging
manifestation of this possibility considering the case of a
Technicolour-type model with a couple of strong vector resonances. The
full details of this model have been already discussed in two previous
references  \cite{strong}, \cite{Zsub},
and we shall not repeat them here. The only thing
that we will show are the characteristic regions of the model, that is
essentially descriable by two parameters. As one can see in Figs.1,2 the
visible regions (where the size of the effect is larger than that of
the realistic experimental error \cite{LEP})
of the AGC and of the TC models are indeed well separated, and
no confusion between these two models would possibly arise.\par
Having illustrated, we hope in a clear way, the main features of our
approach for a specific type of (almost) universal new physics effects,
we shall devote the next and last section to the discussion of a
"typically" non universal kind of effects, generated by the presence of
one extra (and of the most general type) Z.

\section{A Model with one General Extra Z}

As a possibly rewarding unconventional application of our method, we
illustrate the treatment of a model where one extra $Z$ (generically
denoted $Z'$), with the most general type of vector and axial couplings
to leptons and quarks, is supposed to exist. All the popular
"canonical" models ($E_6$, $LR$ symmetry, composite models,...) will be
then recovered by adjusting the couplings to the corresponding
values.\par
The effect of a heavy $Z'$, of a mass not smaller than $\simeq 400-500
GeV$, as \underline{suggested} from the available CDF limits
\cite{CDFZ},is usually treated
at "$Z'$-tree level" i.e. only adding to the full
amplitude the graph with the $Z'$ exchange, where both its couplings to
fermions and its mass are identified with the physical ones. This leads
to a modification of the Born amplitude of the following form:

\bq \A^{\gamma, Z, Z'}_{lf} =  \A^{\gamma, Z}_{lf} +
{i\over q^2-M^2_{Z'}} v_{\mu,l}^{(Z')} v_{\mu,f}^{(Z')}\eq

\noindent
with the general $Z'ff$ couplings

\bq v_{\mu,f}^{(Z')} =
({e\over2c_1s_1})\bar u_f\gamma_{\mu}(g^{,}_{Vf}
-\gamma^5 g^{,}_{Af})v_f \eq

{}From a formal point of view, that will be particularly suited for our
approach, it is possible to rewrite the $Z'$ effect as a modification
of our "generalized" subtracted corrections. This effect, that would
correspond exactly to a "box-type" modification of completely non
universal type, can be described in the following way:

\bq  \tilde{\Delta}^{(lf)(Z')}\alpha(q^2) =
{q^2\over q^2-M^2_{Z'}}({1\over 4s^2_1c^2_1})({g_{Vl}g_{Vf}\over
Q_lQ_f})
 [(\xi_{Vl}-\xi_{Al})(\xi_{Vf}-\xi_{Af})]
\eq

\bq R^{(lf)(Z')}(q^2) = -({q^2-M^2_{Z}
\over q^2-M^2_{Z'}})
\xi_{Al}\xi_{Af} \eq

\bq V^{(lf)(Z')}_{\gamma Z}(q^2) =
-({q^2-M^2_{Z}\over q^2-M^2_{Z'}})({g_{Vl}\over2s_1c_1Q_l})
\xi_{Af}(\xi_{Vl}-\xi_{Al}) \eq

\bq V^{(lf)(Z')}_{Z\gamma}(q^2) =
-({q^2-M^2_{Z}\over q^2-M^2_{Z'}})({g_{Vf}\over2s_1c_1Q_f})
\xi_{Al}(\xi_{Vf}-\xi_{Af}) \eq

\noindent
where we have used the definitions:

\bq \xi_{Vl,f} \equiv {g'_{Vl,f}\over g_{Vl,f}}   \eq

\bq \xi_{Al,f} \equiv {g'_{Al,f}\over g_{Al,f}}   \eq

\bq  g_{Al,f} \equiv I^{3L}_{l,f} \eq

\bq  g_{Vl,f} \equiv I^{3L}_{l,f}-2Q_{l,f}s^2_1 \eq

One sees from eqs.(90)-(92) that the most general $Z'$ effect at
$e^+e^-$ colliders is parametrizable via \underline{six} independent
effective couplings, that could be chosen as e.g.
$\xi_{V,A,i}{M_Z\over\sqrt{M^2_{Z'}-q^2}}$  ($i=l, u, d$).
Therefore, with one experiment at fixed $q^2$ it would never be
possible to disentangle $\xi_{V,A}$ from $M_{Z'}$, so that the normal
attitude would be to derive (in case of negative searches) bounds for
$M_{Z'}$ for given $\xi_{V,A}$. In fact, this will be done in another
specific dedicated paper in preparation. Here we want to show that, in
full analogy with the final example of the previous section, it would
be possible to draw a region in a 3-dimensional space of observables
that would be typical of the \underline{most general} $Z'$. To achieve
this goal, one must necessarily choose three purely leptonic
onservables.\par
At LEP2, this might be obtained by combining the measurements of
$\sigma_{\mu}$ and $A_{FB,\mu}$ with that of the final $\tau$
polarization. At NLC, the role of the final $\tau$ polarization would
be played by the (theoretically equivalent) longitudinal polarization
asymmetry for leptons. The general $Z'$ contribution to these
quantities will actually take the form of eq.(71)-(73) with
$\tilde{\Delta}^{(AGC)}\alpha(q^2)$, $R^{(AGC)}(q^2)$, $V^{(AGC)}(q^2)$
respectively replaced by
$\tilde{\Delta}^{(Z')}$, $R^{(Z')}(q^2)$ and $V^{(Z')}(q^2)$
given in eq.
(93)-(95) for $f=l$.\par
Eliminating the two effective leptonic parameters gives then rise to a
relationship between the shifts of $\sigma_{\mu}$, $A_{FB,\mu}$ and
$A_{\tau}$ that would lead, at LEP2 energies, to a certain
3-dimensional region characteristic of this model and represented in
Fig.3 (we assumed the same experimental errors as in the previous
figures). Note that, with this procedures, all residual "intrinsic" Z'
ambiguities e.g. in the normalization of $g'_V$, $g'_A$ disappear.\par
 A warning is necessary at this point since this Figure, as well
as the previous ones, have been drawn in  "first approximation" i.e.
without calculating the fully QED convoluted effects (this is, in fact,
in preparation at the moment). We can, though, claim that, as a general
feature of such more realistic calculations, the "first approximation"
results are quite reasonably reproduced provided that a suitable cut is
enforced on the hard photon spectrum. In this spirit, we believe that
it makes sense to compare Fig.3 for the $Z'$ model with the
corresponding Fig.1 for the AGC and TC models and conclude that, at
least in this orientative picture, the three regions corresponding to
these theoretically "orthogonal" models are completely (i.e. in the
physically reasonable region where a statistical meaning can be
attributed to the signal) separated.

\section{Concluding remarks}
We have shown in this paper that the calculation of new physics effects
in a general four-fermion process is facilitated if the
procedure of "trading" $G_{\mu}$ by quantities measured on Z peak is
generalized from the case of final leptonic states to that of final
hadronic states. The new relevant quantities that enter the modified
Born Approximation are the Z hadronic widths $\Gamma_5$ and $\Gamma_b$
and, to a much smaller extent the charm width $\Gamma_c$ and the two
forward-backward asymmetries $A_{FB,c}$, $A_{FB,b}$ on Z resonance, if
we only consider the measurements of $\sigma_5$, $\sigma_b$ and
$A_{FB,b}$ at variable $q^2$. We want to conclude this paper by making
this statement more quantitative.\par
Consider $\sigma_5$ first. Here the leading terms at Born level are the
pure photon and the pure $Z$ contributions. In our modified expression,
the only Born term that changes is that corresponding to $Z$ exchange,
whose numerical weight is roughly of the same size as that of the
photon. The net effect of the change is that of replacing here
$G^2_{\mu}$ by the product of $\Gamma_l$ and $\Gamma_5$.
The corresponding relative experimental error thus introduced is a
fraction of a percent \cite{LEP}, much below the experimental
reach at any future $e^+e^-$ collider. The same conclusion applies
to the term $N_f=3(1+{\alpha_S(M^2_Z)\over\pi})$ that
divides $\Gamma_5$ and that generates an error of a few per mille
at most. Note that the same relative error will affect the
contribution that we called "QCD", since $\alpha_s(q^2)$ should be
known with the same accuracy as  $\alpha_s(M^2_Z)$.
The remaining new input
quantities that enter $\sigma_5$ are $\Gamma_c$ and
$s^2_{b,c}(M^2_Z)$ defined by eq.(33). But even without discussing this
point in full detail, as one could easily do, one sees immediately that
these new parameters only contribute to the interference $\gamma-Z$
. The latter is, already at the starting Born level, completely
negligible with respect to the dominant pure photon and $Z$ ones.
Therefore, a discussion on the effect of "small" changes in this term
is, indeed, completely academic and we shall not give it here.\par
In the case of $\sigma_b$, the same situation is almost identically
reproduced, with the only replacement of $\Gamma_5$ by $\Gamma_b$ in
the $Z$ Born expression. The error on $\Gamma_b$ is in fact slightly
larger, of a relative one percent \cite{LEP}, but also the experimental
accuracy for $\sigma_b$ will be certainly larger than one percent, and
the same conclusions as in the case of $\sigma_5$ still apply.\par
The last case to be discussed is that of $A_{FB,b}$. Here the situation
is quite different since the $\gamma-Z$ term is now largely dominating.
This term contains $\Gamma_l$, $\Gamma_b$, that will introduce errors
of negligible size (i.e. at the relative level of less than one
percent) and a term containing $s^2_b(M^2_Z)$ as one sees from
eq.(67). In fact, the relevant quantity to be considered is

\bq {1\over (1+\tilde{v}^2_b)^{1/2}} \eq

\noindent
that is directly related to the forward-backward asymmetry on $Z$
resonance  $A_{FB,b}(M^2_Z)$ \cite{LEP}. From the 4 percent
uncertainty on this quantity
given in ref.\cite{LEP} one can derive the relative error on the
term in eq.(94) that generates a 3 percent uncertainty on the
prediction for $A_{FB,b}$. This is
also weaker than the experimental uncertainty expected at LEP2.\par

In conclusion, all the replacements in the Born approximation are
completely harmless for the considered process. Therefore, the gain
that we obtained in the corresponding simplifications of the
"subtracted" corrections seems to us rather remarkable.
We would say that the full and rigorous exploitation of the high
precision measurements of electroweak physics at $q^2=M^2_Z$ allows
to perform calculations of virtual new physics effects at LEP2 (and,
possibly, at NLC) in a way that seems to us simpler and cleaner than
the conventional one where $G_{\mu}$, the high precision electroweak
measurement at $q^2=0$, is used. We are now in
the process of applying the method to other possibly interesting models
of new physics for which calculations of virtual effects might be
relevant at future $e^+e^-$ colliders.\par

\vspace{0.5cm}

{\bf \underline{Acknowledgements}}\par
We thank Jacques Layssac for his help in the preparation of the figures.

\newpage

\centerline { {\bf Figure Captions }}

\vspace{0.5cm}

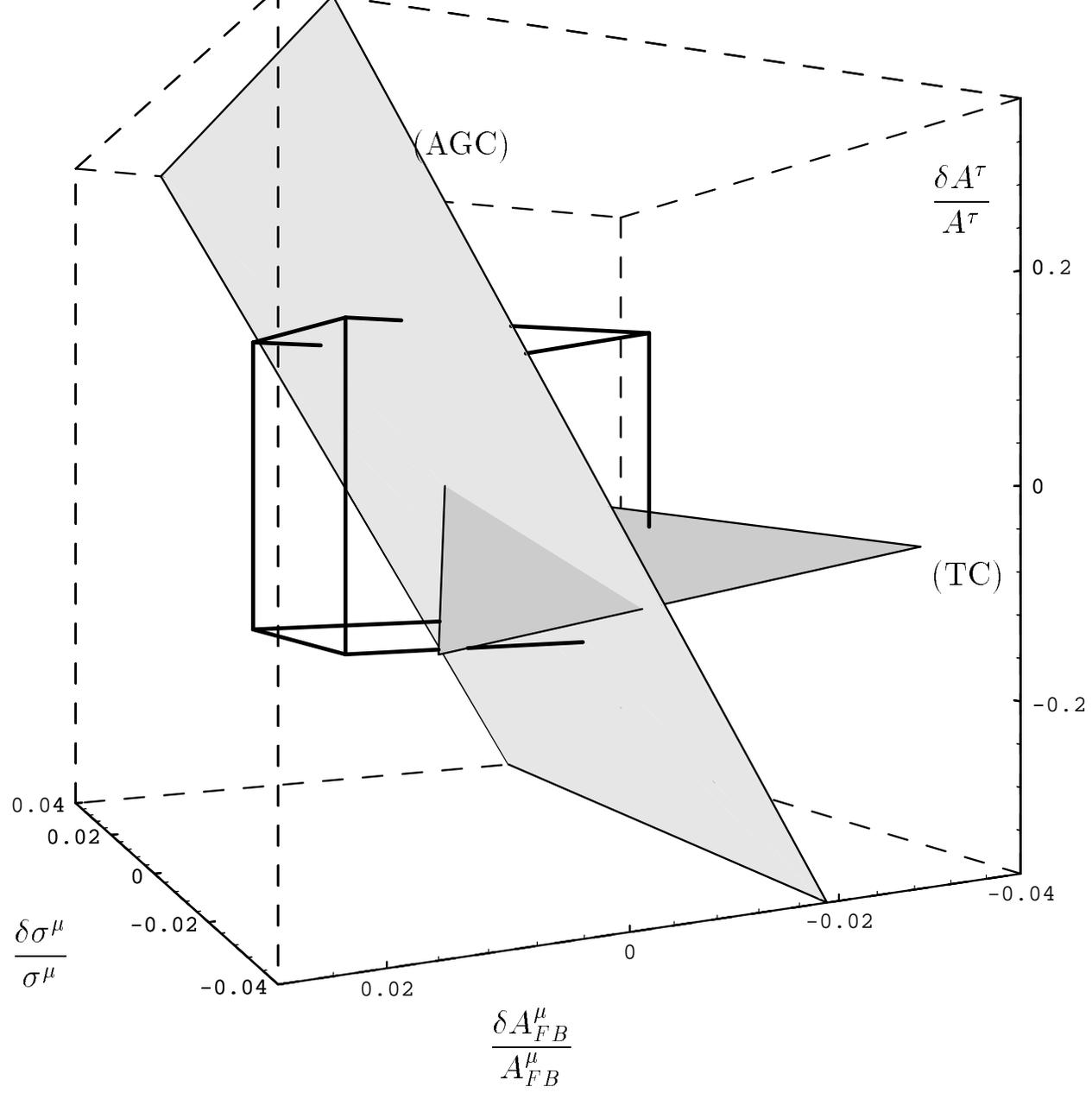
{\bf Fig.1} Trajectories in the 3-dimensional
space of relative departures from SM for
leptonic observables
$\sigma_{\mu}$, $A_{FB,\mu}$, $A_{\tau}$ at a LEP2 energy of 175 GeV
for AGC
models and TC models. The box represents the unobservable domain
corresponding to a relative accuracy of 1.5 percent for
$\sigma_{\mu}$, $A_{FB,\mu}$ and 15 percent for $A_{\tau}$.\\

\vspace{0.5cm}
 {\bf Fig.2} Trajectories in the 3-dimensional
space of relative departures from SM for
leptonic and hadronic observables
$\sigma_{\mu}$, $A_{FB,\mu}$, $R_5$  at a LEP2 energy of 175 GeV
for AGC
models and TC models. The box represents the unobservable domain
corresponding to a relative accuracy of 1.5 percent for
all three observables.\\
\vspace{0.5cm}

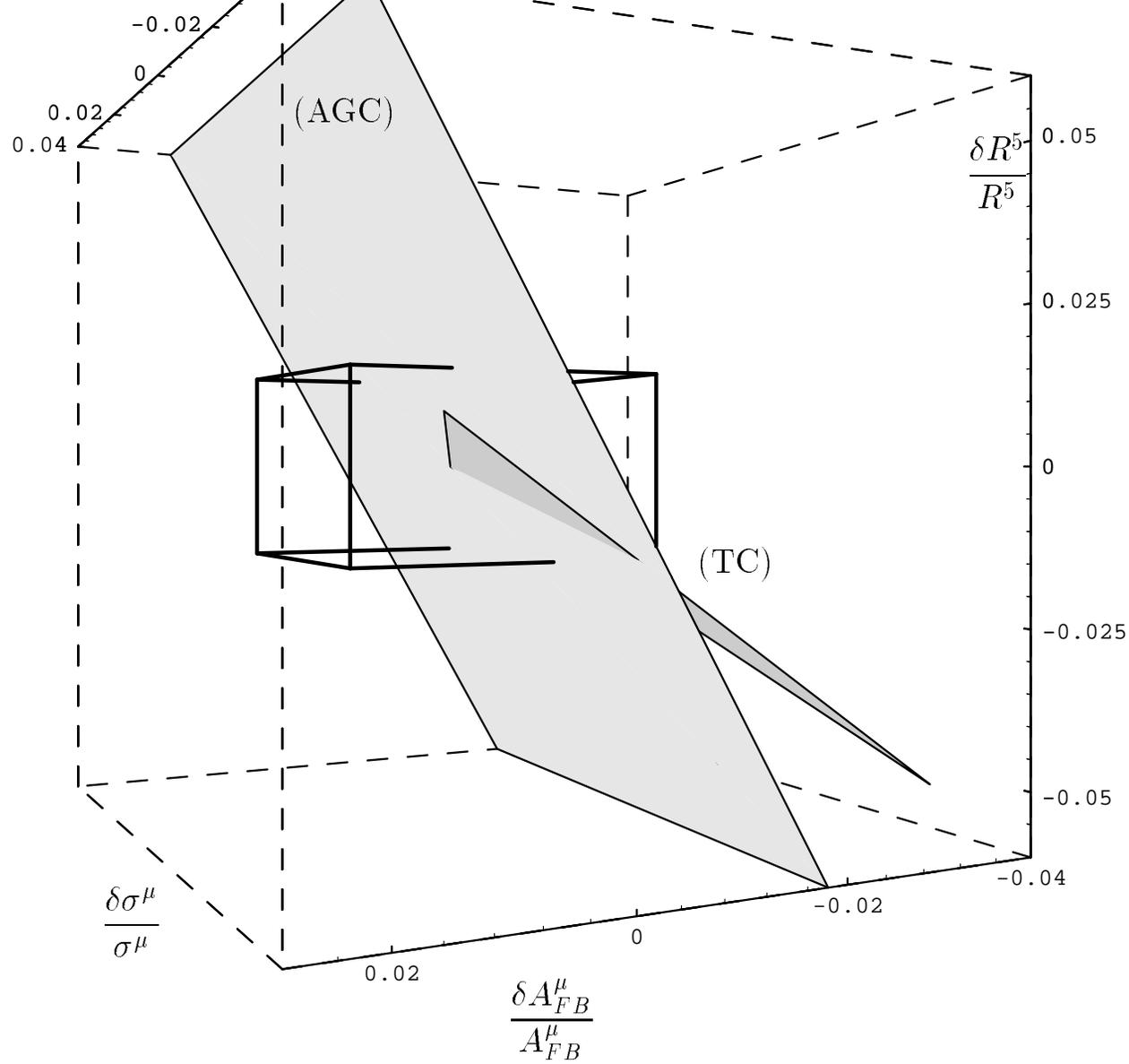
{\bf Fig.3} Trajectories in the 3-dimensional
space of relative departures from SM for
leptonic observables
$\sigma_{\mu}$, $A_{FB,\mu}$, $A_{\tau}$  at a LEP2 energy of 175 GeV
for general $Z'$
models. The box has the same meaning as in Fig.1.

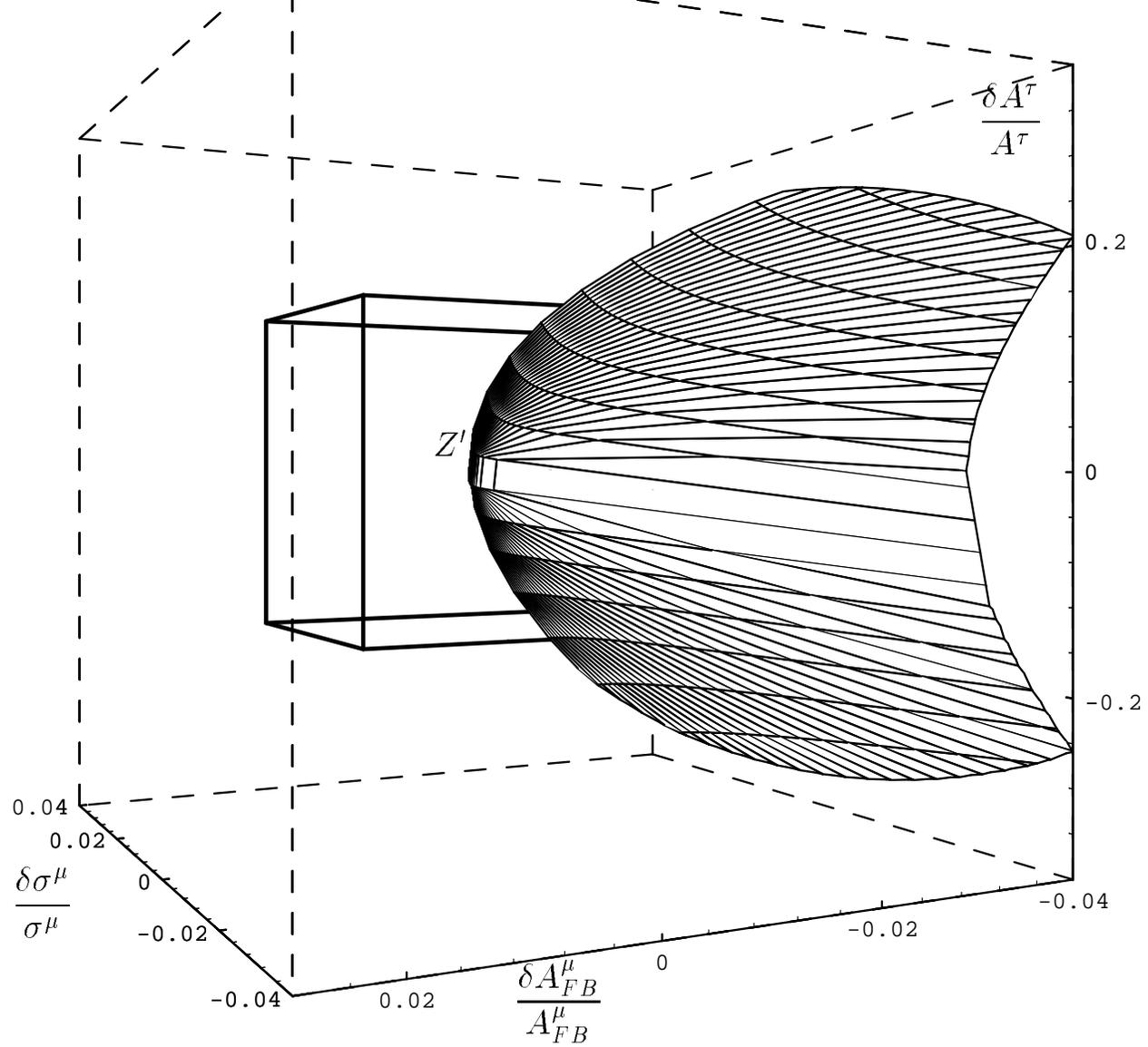
\end{document}